\documentstyle[12pt]{article} 
\begin{document} 

{ \pagestyle{empty} 
\vskip 3cm 
\centerline{\Large \bf Weber-Fechner's Law and Demand Function}

\vskip 3cm

\centerline{K. Shigemoto\footnote{E-mail address:
shigemot@tezukayama-u.ac.jp}}
\centerline {{\it Department of Economics, Tezukayama University}}
\centerline {{\it Tezukayama 7, Nara 631, Japan}}

\vspace{3cm}

\centerline{\bf Abstract} \vspace{5mm} 

We apply the Weber-Fechner's law, which represents the relation 
between the magnitude of physical stimulus and the magnitude of 
psychological sense in human being, to the utility function.
We conclude that the utility function of $n$-types of goods is 
of separable type $u(x_1,x_2,\cdots, x_n)=u_1(x_1)+u_2(x_2)+\cdots 
+u_n(x_n)$, which gives the relation of the demand function in 
the form 
$p_i=d u_i/d x_i$. 
The explicit quantitative form of each utility function, which is suggested 
by the Weber-Fechner's law,  becomes 
$u_i(x_i)=A_i \log (x_i/x^{(0)}_i)$. Then we obtain each demand function 
in the familiar form $p_i=A_i/x_i$.

\newpage
}
\noindent {\large \bf \S 1.\ \ Introduction} 

\vspace{2mm} 

Nowadays it is widely said that the 20-th century {\it was}
the century of physics but the 21st century {\it is} the 
century of biology. 
Here and there, the new trend of physics gradually 
appears, and pioneers in such new trend physics consider that 
the physics should treat not only the science of the nature 
but also the science of human being\cite{Penrose,Wolfram}. 
As a result of the advance of the study of the complex system, 
we may realize the dream to understand the origin of life 
in near future.  Today, the progress in biophysics, biochemistry,
gene technology, brain physics {\it etc.} is surprising. 

In near future, phenomena connected with human being 
such as behavioral psychological phenomena will be 
understand from physics. 
The important part of economics depends on the knowledge 
of behavioral psychology, especially behavioral psychology 
concerning about money.
In this context, we consider economics as the application 
of behavioral psychology.
In micro economics, the utility function is one of the most
important things, and the functional form of this utility 
function must be determined from the knowledge of  
behavioral psychology. 
In macro economics, functional dependence of various macro 
functions such as the consumption function, the saving function
and the investment function {\it etc.} must be determined from
 the knowledge of behavioral psychology.
For example, the consumption function $C$ is often 
assumed as $C=C(Y-T)=C_0+c(Y-T)$ where 
$Y$: national income; $T$: tax; $c$: marginal propensity 
of consume \cite{Samuelson}.
This is considered as the law of behavioral psychology. 
This law is quite simple but universal and powerful, which 
governs the behavior of human being. While, the functional form
of the utility function is still in qualitative level as 
people use this or that functional form without any scientific
reason. 

The most important law in behavioral psychology is 
Weber-Fechner's law\cite{Weber,Fechner},  which represents 
the relation 
between the magnitude of physical stimulus and the 
magnitude of psychological sense in human being. 
In previous paper, we apply this Weber-Fechner's law
to the utility function, and give the explicit quantitative 
functional form for the utility function\cite{Shigemoto}.
In this paper, we give the explicit quantitative functional 
form for the demand function by using the utility 
function proposed in the previous paper.
We also give the interpretation of the relation of this demand 
function from the view point of consumer's surplus.

\vspace{8mm}

\noindent {\large \bf \S 2.  Demand function
from the Weber-Fechner's law}
\vspace{2mm}

The well-known and quite important law in behavioral 
psychology is the Weber-Fechner's law\cite{Weber,Fechner}.
Modification of the Weber-Fechner's law is known as 
the Stevens' law, which is successful in the phenomena
of sound\cite{Stevens}. This Weber-Fechner's law is 
expressed as 
" the magnitude of psychological sense
is proportional to the logarithm of the 
magnitude of physical stimulus."
More precisely, we denote the magnitude of 
psychological sense as $u$ and 
the magnitude of physical stimulus
as $x$, then we can express the Weber-Fechner's 
law in the following form
 
\begin{eqnarray}
u(x)=\left\{ \begin{array}{ll}
             A \log{\left( x/x^{(0)} \right)} & 
                    \mbox{if $x \ge x^{(0)}$ } \\
             0   & \mbox{ $x<x^{(0)}$ } \end{array} 
     \right.  ,
\label{e1}
\end{eqnarray}

\noindent
where $A$ is constant and $x^{(0)}$ is the threshold 
of the  magnitude of the physical stimulus. 
In the following, we consider only in the region 
where the magnitude of the physical stimulus is 
above the threshold value.

We apply this Weber-Fechner's formula to the 
utility function\cite{Shigemoto}.
First, we assume that there is only one type of goods, 
then the Weber-Fechner's law suggests that the utility 
function $u_1$ is given as the function of
the quantity $x_1$ of this goods in the form 
\begin{eqnarray}
 u_1(x_1)=A_1 \log(x_1/x^{(0)}_1)  .
\label{e2}
\end{eqnarray}
\noindent
Next we assume that there are two types of goods, 
then each utility function $u_1$ and $u_2$ is
given as $ u_1(x_1)=A_1 \log(x_1/x^{(0)}_1)$ and
$ u_2(x_2)=A_2 \log(x_2/x^{(0)}_21)$ as the function
of each quantity $x_1$ and $x_2$. It is the natural 
assumption that the total utility function is just the sum 
of each utility function, as each utility obtained by getting 
each goods is independent. 
Then the total utility function in this case is given by 

\begin{eqnarray}
 u(x_1,x_2)&=&A_1 \log(x_1/x^{(0)}_1)+A_2 \log(x_2/x^{(0)}_2)
\label{e3}\\
        &=&\log \left( (x_1/x^{(0)}_1)^{A_1} \times 
     (x_2/x^{(0)}_2)^{A_2} \right).
\label{e4}
\end{eqnarray}
\noindent
The second expression in the above means that the total 
utility function is 
the logarithm of the Cobb-Douglas type function.
We can generalize this analysis to $n$-types of goods,
and the total utility function is given by 

\begin{eqnarray}
&&u(x_1,x_2,\cdots,x_n)=u_1(x_1)+u_2(x_2)+\cdots+u_n(x_n),
\label{e5} \\
&&{\rm where}\nonumber\\
&&u_i(x_i)=A_i  \log(x_i/x^{(0)}_i).  
\label{e6}
\end{eqnarray}

Next, we derive the demand function by using this utility 
function. 

We assume that each price $p_i$ of $i$-th goods is given.
According to the standard analysis to maximize the 
utility function under budget constraints, we have 

\begin{eqnarray}
\frac{ \partial u/ \partial x_1 }{p_1} =
\frac{ \partial u/ \partial x_2 }{p_2} =
\cdots
=\frac{ \partial u/ \partial x_n }{p_n} 
=k(x_1,x_2,\dots,x_n) . 
\label{e7}
\end{eqnarray}
Using the utility function of separable type Eq(\ref{e5}),
we have 

\begin{eqnarray}
\frac{ du_1(x_1)/ dx_1 }{p_1} =
\frac{ du_2(x_2)/ dx_2 }{p_2} =
\cdots
=\frac{ du_n(x_n)/ dx_n }{p_n} =k(x_1,x_2,\dots,x_n).
\label{e8}
\end{eqnarray}
\noindent
The above relation is satisfied for any number $n$ of types 
of goods, 
which means that $k(x_1,x_2,\cdots,x_n)=$ const.,
because $k(x_1,x_2,\cdots,x_n)$ is 
independent of the type of goods.
Normalizing the utility function, we can choose $k=1$ 
and obtain the relation of the demand function in the form 

\begin{eqnarray}
p_i=\frac{du_i(x_i)}{dx_i} \quad  (i=1,2,\dots,n) ,
\label{e9}
\end{eqnarray}
\noindent 
which is correct for the general utility function of the
separable type. The above argument is interpreted as 
follows. Suppose we consider the $(n+1)$-th goods as money 
itself, then the $(n+1)$-th utility function $u_{n+1}(x_{n+1})$,
which utility is measured by the unit of money, is given by 
$u_{n+1}(x_{n+1})=p_{n+1} x_{n+1}$. Then 
$\displaystyle{\frac{du_{n+1}(x_{n+1})/ dx_{n+1}}{p_{n+1}}=1}$, 
which gives $k=1$. 

Further, we use the explicit form Eq.(\ref{e6}) of the utility 
function, which is suggested by the Weber-Fechner's law, and 
we have the explicit quantitative demand function in the 
familiar form 

\begin{eqnarray}
p_i=\frac{A_i}{x_i} \quad  (i=1,2,\dots,n) .
\label{e10}
\end{eqnarray}
In this way, applying the Weber-Fechner's law to each utility 
function and assuming that total utility is just the sum of 
the utility of each goods, we obtain the quantitative demand 
function in the familiar form.

\vspace{8mm}

\noindent {\large \bf \S 3.\ \ Demand 
function from the consumer's surplus} 

\vspace{2mm} 

We derive the relation of the demand function Eq.(\ref{e9}) 
by using the utility function
from the view point of the consumer's surplus. 
Consumer's surplus is defined in the following way 

\begin{eqnarray}
 { \rm  (Consumer's \ surplus)}
 =\int_{0}^{x} p(x')dx'-px ,
 \label{e11}
\end{eqnarray}
\noindent
where $p$ and $x$ is the price and the quantity of goods.
We define the concept of "consumer's profit" $\rho(x)$ as 
$\rho(x)=u(x)-p x$, that is, the "consumer's profit" means 
the profit of money in mind minus the loss of money paid 
to obtain goods.  
Then the concept of the consumer's surplus 
is equal to the "consumer's profit", that is, 
(consumer's surplus)=("consumer's profit") 
$=\rho(x)=u(x)-p x$. 
This gives the relation
\begin{eqnarray}
 \int_{0}^{x} p(x')dx'=u(x) , 
 \label{e12}
\end{eqnarray}
\noindent
where we assume $u(0)=0$ as the utility function should 
have this property. 
Then we again obtain the relation of the demand function 
in the same form as Eq.(\ref{e9}) 

\begin{eqnarray}
p=\frac{du(x)}{dx} ,
\label{e13}
\end{eqnarray}
\noindent
from the view point of the consumer's surplus. 

We define the "producer's profit" $\pi(x)$ as 
$\pi(x)=px-c(x)$ where $c(x)$ is the cost 
function. As is well-known, the supply function 
is derived by maximizing this  "producer's profit".
Then the relation of the supply function is given by 

\begin{eqnarray}
{\rm \underline{Supply \ function}:} \quad
\frac{d\pi(x)}{dx}=p-\frac{dc(x)}{dx}=0 .
\label{e14}
\end{eqnarray}

Similarly, we can derive the demand function 
by maximizing the "consumer's profit ", $\rho(x)=u(x)-px$.
Then the relation of the demand function is given in 
the same form as Eq.(\ref{e9})  

\begin{eqnarray}
{\rm \underline{Demand \ function}:} \quad
\frac{d\rho(x)}{dx}=\frac{du(x)}{dx}-p=0  .
\label{e15}
\end{eqnarray}

In this way, the relation of the demand function is 
derived just in the same 
way as the the relation of the supply function from 
the view point of the 
consumer's surplus and we again obtain the same 
relation Eq.(\ref{e9}) for the demand function.

If we assume $u(0) \ne 0$, the "consumer's profit" differs from 
the consumer's surplus by only the constant value $u(0)$. Then we can 
say that the demand function is determined  by maximizing the 
"consumer's profit" or by maximizing the consumer's surplus. 
Similarly, the  "producer's profit" differs from 
the producer's surplus by only the constant value $c(0)$. Then we can 
say that the supply function is determined  by maximizing the 
"producer's profit" or by maximizing the producer's surplus.

\vspace{8mm}

\noindent {\large \bf \S 4. Summary and discussion}

We try to reconstruct economics as the application of 
behavioral psychology. The most important law in 
behavioral psychology is the Weber-Fechner's law, 
which represents the relation between the magnitude 
of physical stimulus and the magnitude of 
psychological sense in human being.
While the utility function is one of the most important 
thing in micro economics. In this paper, we apply the 
Weber-Fechner's law to determine the explicit quantitative 
utility function. 
Then utility function becomes in the form of 
the logarithm of the Cobb-Douglas type function. 
We conclude that $i)$ the utility function of $n$-types 
of goods are separable 
$u(x_1,x_2,\cdots, x_n)=u_1(x_1)+u_2(x_2)+\cdots 
+u_n(x_n)$, $ii)$each utility function becomes
$u_i(x_i)=A_i \log (x_i/x^{(0)}_i)$. From the property $i)$,
we have the relation of the demand function 
in the form $p_i=d u_i/d x_i$. Using also the property 
$ii)$, which is suggested by the Weber-Fechner's law,
we have the explicit quantitative demand function 
in the familiar form $p_i=A_i/x_i$. We also derive the 
relation of the demand 
function from the view point of the consumer's surplus and 
obtain the same relation $p_i=d u_i/d x_i$ for the demand 
function. 

In this way, as the application of the Weber-Fechner's law 
to the utility function, we obtain the quantitative demand 
function in the familiar form.
Our analysis is the zero-th order approximation for the demand 
function. If we modify the Weber-Fechner's law into Stevens'
law, we obtain the more complicated demand function.

\vspace{8mm}

\newpage


\end{document}